\documentclass[a4paper, 12pt]{article}
\usepackage{amsmath}
\usepackage{graphicx}
\usepackage{amsfonts}
\usepackage{amssymb}
\def \det{\mathrm{det}}

\newtheorem{Lemma}{Lemma}
\newtheorem{Theorem}{Theorem}

\title{$N$-Dark-Dark Solitons in the Generally Coupled Nonlinear Schr\"{o}dinger Equations}
\author{Yasuhiro Ohta$^1$\footnote{Email: ohta@math.kobe-u.ac.jp}, ~~
Deng-Shan Wang$^2$\footnote{Email: wangdsh1980@yahoo.com.cn},~~
Jianke Yang$^3$\footnote{E-mail: jyang@cems.uvm.edu, corresponding author.} \\
{\small\it $^1$Department of Mathematics, Kobe University, Rokko, Kobe 657-8501, Japan} \\
{\small\it $^2$CEMA, Central University of Finance and Economics, Beijing, 100081, China}\\
  {\small\it $^3$Department of Mathematics and Statistics, University of Vermont, Burlington, VT $05401$, U.S.A.}}

\date{July 14, 2010}
\begin{document}

\maketitle
\begin{abstract}
$N$-dark-dark solitons in the generally coupled integrable NLS
equations are derived by the KP-hierarchy reduction method. These
solitons exist when nonlinearities are all defocusing, or both
focusing and defocusing nonlinearities are mixed. When these
solitons collide with each other, energies in both components of the
solitons completely transmit through. This behavior contrasts
collisions of bright-bright solitons in similar systems, where
polarization rotation and soliton reflection can take place. It is
also shown that in the mixed-nonlinearity case, two dark-dark
solitons can form a stationary bound state.

\end{abstract}

\section{Introduction}

In studies of nonlinear wave dynamics in physical systems, nonlinear
Schr\"{o}dinger (NLS)-type equations play a prominent role. It is
known that a weakly nonlinear one-dimensional wave packet in a
generic physical system is governed by the NLS equation
\cite{Benney}. Hence this equation appears frequently in nonlinear
optics and water waves \cite{Agrawal_book, Hasegawa_book,
Ablowitz_Segur}. Recently, it has been shown that the nonlinear
interaction of atoms in Bose-Einstein condensates is governed by a
NLS-type equation as well (called Gross-Pitaevskii equation in the
literature) \cite{Dalfovo_1999}. In these physical systems, the
nonlinearity can be focusing or defocusing (i.e., the nonlinear
coefficient can be positive or negative), depending on the physical
situations \cite{Ablowitz_Segur} or the types of atoms in
Bose-Einstein condensates \cite{Dalfovo_1999}. When two wave packets
in a physical system or two types of atoms in Bose-Einstein
condensates interact with each other, their interaction then is
governed by two coupled NLS equations \cite{Agrawal_book,
Hasegawa_book,Ho_Shenoy,Pu_Bigelow_1,Pu_Bigelow_2,Goldstein_Meystre,Roskes,Menyuk_1987}.
The single NLS equation is exactly integrable
\cite{Zakharov_Shabat}. It admits bright solitons in the focusing
case, and dark solitons in the defocusing case. Its bright
$N$-soliton solutions were given in \cite{Zakharov_Shabat}, and its
dark $N$-soliton solutions can be found in \cite{Faddeev_book}. The
coupled NLS equations are also integrable when the nonlinear
coefficients have the same magnitudes
\cite{Manakov,Zakharov1982,Wang2010}. In these integrable cases, if
all nonlinear terms are of focusing type (i.e., the nonlinear
coefficients are all positive), the coupled NLS equations are the
focusing Manakov model which admits bright-bright solitons
\cite{Manakov}. If all nonlinear terms are of defocusing type (i.e.,
the nonlinear coefficients are all negative), the coupled NLS
equations are the defocusing Manakov model which admits bright-dark
and dark-dark solitons \cite{Sheppard_Kivshar_1997,RL,Ablowitz1}. If
the focusing and defocusing nonlinearities are mixed (i.e., the
nonlinear coefficients have opposite signs), these coupled NLS
equations admit bright-bright solitons \cite{Wang2010,KLTA} and
bright-dark solitons \cite{VKL_2008}. Existence of dark-dark
solitons in this mixed case has not been investigated yet.

Soliton interaction in these integrable generally coupled NLS
equations is a fascinating subject. In the focusing Manakov model,
an interesting phenomenon is that bright solitons change their
polarizations (i.e. relative energy distributions among the two
components) after collision \cite{Manakov}. In the coupled NLS
equations with mixed nonlinearities, energy can also transfer from
one soliton to another after collision \cite{KLTA}. In addition,
solitons can be reflected off by each other as well \cite{Wang2010}.
In the defocusing Manakov model, two bright-dark solitons can form a
stationary bound state, a phenomenon which does not occur for scalar
bright or dark solitons \cite{Sheppard_Kivshar_1997}. All these
interesting interaction behaviors can be described by multi-soliton
solutions in the underlying integrable system. In the focusing
Manakov model, $N$-bright-bright solitons were derived in
\cite{Manakov} by the inverse scattering transform method. In the
mixed-nonlinearity model, two- and three-bright-bright solitons and
two-bright-dark solitons were derived in \cite{KLTA,VKL_2008} by the
Hirota method, and $N$-bright-bright solitons were derived in
\cite{Wang2010} by the Riemann-Hilbert method. In the defocusing
Manakov model, $N$-bright-dark solitons were derived in
\cite{Sheppard_Kivshar_1997}, and degenerate two-dark-dark solitons
were derived in \cite{RL}, both by the Hirota method.

So far, progress on dark-dark solitons in the integrable generally
coupled NLS equations is very limited. While dark-dark solitons in
the defocusing Manakov model were derived in \cite{RL}, we will show
that their two- and higher-dark-dark solitons are actually
degenerate and reducible to scalar dark solitons. In
\cite{Ablowitz1}, the inverse scattering transform method was
developed for dark solitons in the defocusing Manakov model. But, as
we will show in this paper, their analysis can not yield general
dark-dark solitons either due to their choices of the boundary
conditions. To date, general multi-dark-dark solitons in the coupled
NLS equations have never been reported yet (to our knowledge). As we
will see, these general multi-dark-dark solitons are not easy to
obtain due to non-trivial parameter constraints which must be met.

In this paper, we comprehensively analyze dark-dark solitons and
their dynamics in the generally coupled integrable NLS equations.
First, we show that these coupled NLS equations can be obtained as a
reduction of the Kadomtsev-Petviashvili (KP) hierarchy. Then using
$\tau$-function solutions of the KP hierarchy, we derive the general
$N$-dark-dark solitons in terms of Gram determinants. These
dark-dark solitons exist in both the defocusing Manakov model and
the mixed-nonlinearity model. Recalling that bright-bright solitons
exist in the mixed-nonlinearity model as well \cite{Wang2010,RL}, we
see that the coupled NLS equations with mixed nonlinearities are the
rare integrable systems which admit both bright-bright and dark-dark
solitons. The dark-dark solitons obtained previously in
\cite{RL,Ablowitz1} for the defocusing Manakov model are only
degenerate cases of our general solutions. Next, we analyze
properties of these soliton solutions. For single dark-dark
solitons, we show that the degrees of ``darkness" in their two
components are different in general. When two dark-dark solitons
collide with each other, we show that energies in the two components
of each soliton completely transmit through. This contrasts
collisions of bright-bright solitons in these same equations, where
polarization rotation, power transfer and soliton reflection can
occur \cite{Manakov,Wang2010,KLTA}. Thus dark-dark solitons are much
more robust than bright-bright solitons with regard to collision. In
the case of mixed focusing and defocusing nonlinearities, an
interesting phenomenon is that two dark-dark solitons can form a
stationary bound state. This is the first report of
dark-dark-soliton bound states in integrable systems. However, three
or more dark-dark solitons can not form bound states, as we will
show in this paper.

We should mention that this KP-hierarchy reduction for deriving
soliton solutions in integrable systems was first developed by the
Kyoto school in the 1970s \cite{DKJM}. So far, this
method has been applied to derive bright solitons in many equations
such as NLS, modified KdV, Davey-Stewartson equations \cite{T,Date,O}.
This method has also
been applied to derive $N$-dark solitons in the defocusing NLS
equation \cite{O}. But this reduction for dark-dark solitons in the
generally coupled NLS equations is more subtle and has never been
done before. In this paper, we will derive general $N$-dark-dark
solitons by this KP-hierarchy reduction and the grace of deep use of
determinant expressions. Compared to the inverse scattering
transform method \cite{Ablowitz1} and the Hirota method \cite{RL},
our treatment is much more clean, and the solution formulae much
more elegant and general. Thus, the KP-reduction method has a
distinct advantage in derivations of dark-soliton solutions.

\section{The $N$-dark-dark solitons}
The generally coupled integrable NLS equations we investigate in
this paper are
\begin{equation}
\begin{array}{ll}
iu_t=u_{xx}+(\delta|u|^2+\epsilon|v|^2)u, \\
iv_t=v_{xx}+(\delta|u|^2+\epsilon|v|^2)v,
\end{array} \label{(1.1)}
\end{equation}
where $\delta$ and $\epsilon$ are real coefficients. This system is
integrable \cite{Manakov,Zakharov1982,Wang2010}. Through $u$ and $v$
scalings, the nonlinear coefficients $\delta$ and $\epsilon$ can be
normalized to be $\pm 1$ without loss of generality. When
$\epsilon=\delta=1$, this system is the focusing Manakov model which
supports bright-bright solitons \cite{Manakov}. When
$\epsilon=\delta=-1$, this system is the defocusing Manakov model
which supports bright-dark and dark-dark solitons
\cite{Sheppard_Kivshar_1997,RL,Ablowitz1}. When $\epsilon$ and
$\delta$ have opposite signs, the system exhibits mixed focusing and
defocusing nonlinearities. In this case, these equations support
bright-bright solitons \cite{Wang2010,KLTA}, bright-dark solitons
\cite{VKL_2008}, and dark-dark solitons (as we will see below).

In this section, we derive the general formulae for $N$-dark-dark
solitons in the integrable coupled NLS system \eqref{(1.1)}. The
basic idea is to treat Eq. \eqref{(1.1)} as a reduction of the KP
hierarchy. Then dark solitons in Eq. \eqref{(1.1)} can be obtained
from solutions of the KP hierarchy under this reduction. For this
purpose, let us first review Gram-type solutions for equations in
the KP hierarchy \cite{Hirota,MOS,OHTI}.

\begin{Lemma}
Consider the following equations in the KP hierarchy \cite{JM,DJM}
\begin{equation}
\begin{array}{ll}
(\frac{1}{2}D_xD_r-1)\tau(k)\cdot\tau(k)=-\tau(k+1)\tau(k-1),\\
(D_x^2-D_y+2aD_x)\tau(k+1)\cdot\tau(k)=0,
\end{array} \label{Lemma1-1}
\end{equation}
where $D$ is the Hirota derivative defined by
\begin{equation}
D_x^mD_y^n \hspace{0.03cm} f(x, y)\cdot g(x, y)\equiv
\left(\frac{\partial}{\partial x}-\frac{\partial}{\partial
x'}\right)^m \left(\frac{\partial}{\partial
y}-\frac{\partial}{\partial y'}\right)^n f(x, y) \hspace{0.06cm}
g(x', y')|_{x=x', \hspace{0.08cm} y=y'} \hspace{0.08cm},
\end{equation}
$a$ is a complex constant, $k$ is an integer, and $\tau(k)$ is a
function of three independent variables $(x,y,r)$. The Gram
determinant solution $\tau(k)$ of the above equations is given by
$$
\tau(k)=\det_{1\le i,j\le N}\Big(m_{ij}(k)\Big)
=\Big|m_{ij}(k)\Big|_{1\le i,j\le N},
$$
where the matrix element $m_{ij}(k)$ satisfies
\begin{equation}
\begin{array}{ll}
\partial_xm_{ij}(k)=\varphi_i(k)\psi_j(k),\\
\partial_ym_{ij}(k)=(\partial_x\varphi_i(k))\psi_j(k)
-\varphi_i(k)(\partial_x\psi_j(k)),\\
\partial_rm_{ij}(k)=-\varphi_i(k-1)\psi_j(k+1),\\
m_{ij}(k+1)=m_{ij}(k)+\varphi_i(k)\psi_j(k+1),
\end{array} \label{Lemma1-2}
\end{equation}\\
and $\varphi_i(k)$ and $\psi_j(k)$ are arbitrary functions
satisfying
\begin{equation}
\begin{array}{ll}
\partial_y\varphi_i(k)=\partial_x^2\varphi_i(k),\\
\varphi_i(k+1)=(\partial_x-a)\varphi_i(k),\\
\partial_y\psi_j(k)=-\partial_x^2\psi_j(k),\\
\psi_j(k-1)=-(\partial_x+a)\psi_j(k).
\end{array} \label{Lemma1-3}
\end{equation}
\end{Lemma}

Before proving this lemma, several remarks are in order. The first
equation in \eqref{Lemma1-1} is the bilinear equation for the
two-dimensional Toda lattice (see p.984 of \cite{JM} and p.4130 of
\cite{DJM}), and the second equation in \eqref{Lemma1-1} is the
lowest-degree bilinear equation in the 1st modified KP hierarchy
(see p.996 of \cite{JM}). Since the two-dimensional Toda lattice
hierarchy and modified KP hierarchies are closely related to the
(single-component) KP hierarchy, all these hierarchies will be
called the KP hierarchy in this paper. Regarding the parameter $a$
in the second equation in \eqref{Lemma1-1}, it corresponds to the
wave-number shift $k_0$ in \cite{JM} [see Eq. (10.3) there]. The
bilinear equation with this parameter was not explicitly written
down in \cite{JM}, but can be found in \cite{DJM} [see Eq. (N-3)
there]. This parameter can be formally removed by the Galilean
transformation for $y$ in \eqref{Lemma1-1}. But for our purpose, it
proves to be important to keep this parameter, as it will pave the
way for the introduction of another similar parameter $b$ in Lemma 2
later. In that case, $a$ and $b$ can not be removed simultaneously
by the Galilean transformation, and they are essential for the
construction of non-degenerate dark-dark solitons in the generally
coupled NLS system (\ref{(1.1)}).

\noindent{\bf \em Proof of Lemma 1.} By using \eqref{Lemma1-2} and
\eqref{Lemma1-3}, we can verify that the derivatives and shifts of
the $\tau$ function are expressed by the bordered determinants as
follows
\[ \begin{array}{ll}
\partial_x\tau(k)=\left|\begin {array}{cc}
m_{ij}(k) &\varphi_i(k) \cr -\psi_j(k) &0 \end {array}\right|,\\
\partial_x^2\tau(k)=\left|\begin {array}{cc}
m_{ij}(k) &\partial_x\varphi_i(k) \cr -\psi_j(k) &0 \end
{array}\right| +\left|\begin {array}{cc} m_{ij}(k) &\varphi_i(k) \cr
-\partial_x\psi_j(k) &0 \end {array}\right|,\\
\partial_y\tau(k)=\left|\begin {array}{cc}
m_{ij}(k) &\partial_x\varphi_i(k) \cr -\psi_j(k) &0 \end
{array}\right| -\left|\begin {array}{cc} m_{ij}(k) &\varphi_i(k) \cr
-\partial_x\psi_j(k) &0 \end {array}\right|,\\
\partial_r\tau(k)=\left|\begin {array}{cc}
m_{ij}(k) &\varphi_i(k-1) \cr \psi_j(k+1) &0 \end {array}\right|,\\
(\partial_x\partial_r-1)\tau(k)=\left|\begin {array}{ccc} m_{ij}(k)
&\varphi_i(k-1) &\varphi_i(k) \cr \psi_j(k+1) &0 &-1 \cr
-\psi_j(k) &-1 &0 \end {array}\right|,\\
 \tau(k+1)=\left|\begin {array}{cc}
m_{ij}(k) &\varphi_i(k) \cr -\psi_j(k+1) &1 \end {array}\right|,\\
\tau(k-1)=\left|\begin {array}{cc} m_{ij}(k) &\varphi_i(k-1) \cr
\psi_j(k) &1 \end {array}\right|,\\
 (\partial_x+a)\tau(k+1)=\left|\begin {array}{cc}
m_{ij}(k) &\partial_x\varphi_i(k) \cr -\psi_j(k+1) &a \end {array}\right|,\\
(\partial_x+a)^2\tau(k+1)=\left|\begin {array}{cc} m_{ij}(k)
&\partial_x^2\varphi_i(k) \cr -\psi_j(k+1) &a^2 \end {array}\right|
+\left|\begin {array}{ccc} m_{ij}(k) &\partial_x\varphi_i(k)
&\varphi_i(k) \cr -\psi_j(k+1) &a &1 \cr -\psi_j(k) &0
&0\end {array}\right|,\\
(\partial_y+a^2)\tau(k+1)=\left|\begin {array}{cc} m_{ij}(k)
&\partial_x^2\varphi_i(k) \cr -\psi_j(k+1) &a^2 \end {array}\right|
-\left|\begin {array}{ccc} m_{ij}(k) &\partial_x\varphi_i(k)
&\varphi_i(k) \cr -\psi_j(k+1) &a &1 \cr -\psi_j(k) &0 &0\end
{array}\right|.
\end{array}
\]
Here the bordered determinants are defined as
\[\left|\begin {array}{cc} m_{ij} &\varphi_i \cr -\psi_j &0
\end {array}\right| \equiv
\left|  \begin{array}{ccccc} m_{11} & m_{12} & \cdots & m_{1N} &
\varphi_1  \\
m_{21} & m_{22} & \cdots & m_{2N} & \varphi_2 \\
\vdots & \vdots & \vdots & \vdots & \vdots \\
m_{N1} & m_{N2} & \cdots & m_{NN} & \varphi_N \\
-\psi_1 & -\psi_2 & \cdots & -\psi_N & 0 \end{array}\right|,
\]
and so on. By using the Jacobi formula of determinants
\cite{Hirota}, we obtain
the bilinear equations \eqref{Lemma1-1} from the above expressions. $\Box$\\

Using Lemma 1, we can obtain solutions to a larger class of
equations in the KP hierarchy below.

\begin{Lemma}
Consider the following equations in the KP hierarchy,
\begin{equation}
\begin{array}{ll}
(\frac{1}{2}D_xD_r-1)\tau(k,l)\cdot\tau(k,l)=-\tau(k+1,l)\tau(k-1,l),\\
(D_x^2-D_y+2aD_x)\tau(k+1,l)\cdot\tau(k,l)=0,\\
(\frac{1}{2}D_xD_s-1)\tau(k,l)\cdot\tau(k,l)=-\tau(k,l+1)\tau(k,l-1),\\
(D_x^2-D_y+2bD_x)\tau(k,l+1)\cdot\tau(k,l)=0,
\end{array} \label{(2.1)}
\end{equation}
where $a, b$ are complex constants, $k, l$ are integers, and
$\tau(k,l)$ is a function of four independent variables $(x,y,r,s)$.
The solution $\tau(k,l)$ to these equations is given by the Gram
determinant
\begin{equation}
\tau(k,l)=\det_{1\le i,j\le N}\Big(m_{ij}(k,l)\Big)
=\Big|m_{ij}(k,l)\Big|_{1\le i,j\le N}, \label{(2.2)}
\end{equation}
where the matrix element $m_{ij}(k,l)$ is defined by
\begin{equation}
\begin{array}{ll}
m_{ij}(k,l)=c_{ij}+\frac{1}{p_i+q_j}\varphi_i(k,l)\psi_j(k,l),\\
\varphi_i(k,l)=(p_i-a)^k(p_i-b)^le^{\xi_i},\\
\psi_j(k,l)=(-\frac{1}{q_j+a})^k(-\frac{1}{q_j+b})^le^{\eta_j},
\end{array} \label{(2.3)}
\end{equation}
with
\begin{equation}
\begin{array}{ll}
\xi_i=p_ix+p_i^2y+\frac{1}{p_i-a}r+\frac{1}{p_i-b}s+\xi_{i0},\\
\eta_j=q_jx-q_j^2y+\frac{1}{q_j+a}r+\frac{1}{q_j+b}s+\eta_{j0},
\end{array} \label{(2.4)}
\end{equation}
and $c_{ij}$, $p_i$, $q_j$, $\xi_{i0}$, $\eta_{j0}$ are complex
constants.
\end{Lemma}

It is noted that the system (\ref{(2.1)}) is an expansion of the
previous system (\ref{Lemma1-1}) by adding a new pair of independent
variables $(s,l)$ to the previous pair $(r,k)$.

\noindent {\bf \em Proof.} It is easy to see that functions
$m_{ij}(k,l)$, $\varphi_i(k,l)$ and $\psi_j(k,l)$ satisfy the
following differential and difference rules,
\begin{equation}
\begin{array}{ll}
\partial_xm_{ij}(k,l)=\varphi_i(k,l)\psi_j(k,l),\\
\partial_ym_{ij}(k,l)=(\partial_x\varphi_i(k,l))\psi_j(k,l)
-\varphi_i(k,l)(\partial_x\psi_j(k,l)),\\
\partial_rm_{ij}(k,l)=-\varphi_i(k-1,l)\psi_j(k+1,l),\\
m_{ij}(k+1,l)=m_{ij}(k,l)+\varphi_i(k,l)\psi_j(k+1,l),\\
\partial_y\varphi_i(k,l)=\partial_x^2\varphi_i(k,l),\\
\varphi_i(k+1,l)=(\partial_x-a)\varphi_i(k,l),\\
\partial_y\psi_j(k,l)=-\partial_x^2\psi_j(k,l),\\
\psi_j(k-1,l)=-(\partial_x+a)\psi_j(k,l).
\end{array} \label{(2.5)}
\end{equation}\\
Then from Lemma 1, we can verify the first two bilinear equations in
\eqref{(2.1)}. The other two equations in \eqref{(2.1)} can be
obtained directly by replacing $a$, $k$, $r$ as $b$, $l$, $s$ in Eq.
(\ref{Lemma1-1}) of Lemma 1. $\Box$

Next, we perform a reduction to the bilinear system (\ref{(2.1)}) in
the KP hierarchy. Solutions to the reduced bilinear equations are
given below.

\begin{Theorem}
Assume that $f$ is a real function of real $x$ and $t,$ and
$g,h$ are complex functions of real $x$ and $t,$ then the following
bilinear equations
\begin{equation}
\begin{array}{ll} (D_x^2+\delta|\mu|^2+\epsilon|\nu|^2)f\cdot f=\delta|\mu|^2
g\bar g+\epsilon|\nu|^2 h\bar h,\\
(iD_t+D_x^2+2icD_x)g\cdot f=0,\\
(iD_t+D_x^2+2idD_x)h\cdot f=0,
\end{array} \label{(2.6)}
\end{equation}
where $\delta$, $\epsilon$, $c$ and $d$ are real constants, $\mu$
and $\nu$ are complex constants, and the overbar `$\ \bar{}\ $'
represents complex conjugate, admit the following solutions,
\begin{equation}
\begin{array}{ll}
f=\Big|\delta_{ij}+\frac{1}{p_i+\bar p_j}
e^{\xi_i+\bar{\xi}_j}\Big|,\\
g=\Big|\delta_{ij}+\frac{1}{p_i+\bar p_j}(-\frac{p_i-ic}{\bar
p_j+ic}) e^{\xi_i+\bar\xi_j}\Big|,\\
h=\Big|\delta_{ij}+\frac{1}{p_i+\bar p_j}(-\frac{p_i-id}{\bar
p_j+id}) e^{\xi_i+\bar\xi_j}\Big|,
\end{array} \label{(2.7)}
\end{equation}
where
\begin{equation}
\xi_j=p_jx+ip_j^2t+\xi_{j0},
\end{equation}
$p_j$ are complex constants satisfying the constraint
\begin{equation}  \label{constraint1}
\frac{\delta|\mu|^2}{|p_j-ic|^2}+\frac{\epsilon|\nu|^2}{|p_j-id|^2}=-2,
\end{equation}
and $\xi_{j0}$ are arbitrary complex constants.
\end{Theorem}

\noindent {\bf \em Proof.} In Lemma 2, if one assumes $x,r,s$ are
real, $y,a,b$ are pure imaginary, $k,l$ are integers, and $ q_j=\bar
p_j, \eta_{j0}=\bar\xi_{j0}, c_{ji}=\bar c_{ij},$ then we have
\begin{equation}  \label{conjugation_constraint_0}
\eta_j=\bar\xi_j, \quad  m_{ji}(k,l)=\overline{m_{ij}(-k,-l)}, \quad
\tau(k,l)=\overline{\tau(-k,-l)}.
\end{equation}
Therefore, defining
\begin{equation}
c_{ij}=\delta_{ij}, \quad {\rm Re}(p_i)>0, \quad f=\tau(0,0), \quad
g=\tau(1,0), \quad h=\tau(0,1),
\end{equation}
where $\delta_{ij}$ is 1 when $i=j$ and 0 otherwise, then
\begin{equation}
f=\Big|m_{ij}(0,0)\Big| =\Big|\delta_{ij}+\frac{1}{p_i+\bar
p_j}e^{\xi_i+\bar\xi_j}\Big|, \qquad \bar g=\tau(-1,0), \qquad \bar
h=\tau(0,-1),
\end{equation}
and
\begin{equation}
\begin{array}{ll}
(\frac{1}{2}D_xD_r-1)f\cdot f=-g\bar g,\\
(\frac{1}{2}D_xD_s-1)f\cdot f=-h\bar h,\\
(D_x^2-D_y+2aD_x)g\cdot f=0,\\
 (D_x^2-D_y+2bD_x)h\cdot f=0.
\end{array} \label{(2.8)}
\end{equation}
Under the above reduction, the solution (\ref{(2.2)}) for $\tau$ can
be rewritten as
\begin{eqnarray}
\tau(k,l) & = & \Big|\delta_{ij}+\frac{1}{p_i+\bar p_j}
(-\frac{p_i-a}{\bar p_j+a})^k(-\frac{p_i-b}{\bar p_j+b})^l
e^{\xi_i+\bar\xi_j}\Big|   \nonumber
\\
&=&e^{\xi_1+\cdots+\xi_N+\bar\xi_1+\cdots\bar\xi_N}
\Big|\delta_{ij}e^{-\xi_i-\bar\xi_i}+\frac{1}{p_i+\bar p_j}
(-\frac{p_i-a}{\bar p_j+a})^k(-\frac{p_i-b}{\bar p_j+b})^l\Big|,
\label{(2.800)}
\end{eqnarray}
with
$$
\xi_i+\bar\xi_i=(p_i+\bar p_i)x+(p_i^2-\bar p_i^2)y
+(\frac{1}{p_i-a}+\frac{1}{\bar p_i+a})r
+(\frac{1}{p_i-b}+\frac{1}{\bar p_i+b})s+\xi_{i0}+\bar\xi_{i0}.
$$
Thus if $p_i$ satisfies the constraint
\begin{equation}
\delta|\mu|^2(\frac{1}{p_i-a}+\frac{1}{\bar p_i+a})
+\epsilon|\nu|^2(\frac{1}{p_i-b}+\frac{1}{\bar p_i+b}) =-2(p_i+\bar
p_i), \label{(2.80)}
\end{equation}
i.e.,
\begin{equation}  \label{constraint2}
\frac{\delta|\mu|^2}{(p_i-a)(\bar
p_i+a)}+\frac{\epsilon|\nu|^2}{(p_i-b)(\bar p_i+b)}=-2,
\end{equation}
then from Eqs. (\ref{(2.800)})-(\ref{(2.80)}), one gets
\begin{equation}
(\delta|\mu|^2\partial_r+\epsilon|\nu|^2\partial_s)\tau(k,l)=-2\partial_x\tau(k,l).
\label{(2.81)}
\end{equation}
Using $f=\tau(0,0)$, this equation gives
\begin{equation}
\delta|\mu|^2f_r+\epsilon|\nu|^2f_s=-2f_x.  \label{(2.82)}
\end{equation}
Differentiation of (\ref{(2.82)}) with respect to $x$ gives
\begin{equation}
\delta|\mu|^2f_{xr}+\epsilon|\nu|^2f_{xs}=-2f_{xx}. \label{(2.83)}
\end{equation}
The first two equations of \eqref{(2.8)} are just
\begin{equation}
f_{xr}f-f_xf_r-f^2=-g\bar g,\label{(2.84)}
\end{equation}
\begin{equation}
f_{xs}f-f_xf_s-f^2=-h\bar h\label{(2.85)}.
\end{equation}
So from Eqs. \eqref{(2.82)}-\eqref{(2.85)}, we have
\begin{equation}
2f_{xx}f-2f_x^2+(\delta|\mu|^2+\epsilon|\nu|^2)f^2=\delta|\mu|^2g\bar
g+\epsilon|\nu|^2h\bar h\label{(2.86)},
\end{equation}
which is just
\begin{equation}
(D_x^2+\delta|\mu|^2+\epsilon|\nu|^2)f\cdot f=\delta|\mu|^2 g\bar
g+\epsilon|\nu|^2 h\bar h.
\end{equation}
Finally, denoting
\begin{equation}
y=it, \quad a=ic, \quad b=id,
\end{equation}
with $t$, $c$ and $d$ real, the second and third equations in
(\ref{(2.6)}) and (\ref{(2.7)}) are obtained directly from Lemma 2,
and the constraint (\ref{constraint1}) is obtained directly from Eq.
(\ref{constraint2}). Theorem 1 is then proved. $\Box$

Now we transform the bilinear equations \eqref{(2.6)} in Theorem 1
into a nonlinear form. To do so, we set
\begin{equation}
\tilde{u}=\mu\frac{g}{f},\qquad \tilde{v}=\nu\frac{h}{f},
\label{(2.11)}
\end{equation}
where $f,g,h$ satisfy Eq. (\ref{(2.6)}). From (\ref{(2.11)}), we
have
 $$
(D_tg\cdot f)/f^2=\tilde{u}_t/\mu,  \quad (D_th\cdot
f)/f^2=\tilde{v}_t/\nu,
$$
$$
(D_xg\cdot f)/f^2=\tilde{u}_x/\mu, \quad (D_xh\cdot
f)/f^2=\tilde{v}_x/\nu,
$$
\begin{equation}
(D_x^2g\cdot f)/f^2=\tilde{u}_{xx}/\mu+(\tilde{u}/\mu)(D_x^2f\cdot
f)/f^2,\label{(2.12)}
\end{equation}
$$(D_x^2h\cdot
f)/f^2=\tilde{v}_{xx}/\nu+(\tilde{v}/\nu)(D_x^2f\cdot f)/f^2.
$$
The first bilinear equation in \eqref{(2.6)} is
$$
D_x^2f\cdot f=-(\delta|\mu|^2+\epsilon|\nu|^2)f^2+\delta|\mu|^2
g\bar g+\epsilon|\nu|^2 h\bar h
$$
which can be further rewritten as
\begin{equation}
(D_x^2f\cdot
f)/f^2=-(\delta|\mu|^2+\epsilon|\nu|^2)+\delta|\tilde{u}|^2
+\epsilon|\tilde{v}|^2, \label{(2.13)}
\end{equation}
The second bilinear equation in \eqref{(2.6)} is just
\begin{equation}
\frac{(D_x^2+iD_t+2icD_x)g\cdot f}{f^2}=0. \label{(2.14)}
\end{equation}
Substituting \eqref{(2.12)} into \eqref{(2.14)}, we have
\begin{equation}
i\tilde{u}_t+\tilde{u}_{xx}+\tilde{u}(D_x^2f\cdot
f)/f^2+2ic\tilde{u}_x=0. \label{(2.15)}
\end{equation}
In the same way, from the third bilinear equation in \eqref{(2.6)}
we have
\begin{equation}
i\tilde{v}_t+\tilde{v}_{xx}+\tilde{v}(D_x^2f\cdot
f)/f^2+2id\tilde{v}_x=0. \label{(2.16)}
\end{equation}
 Substituting \eqref{(2.13)} into \eqref{(2.15)} and
\eqref{(2.16)}, we get
\begin{equation}
\begin{array}{ll}
i\tilde{u}_t+2ic\tilde{u}_x+\tilde{u}_{xx}+\tilde{u}[-\delta|\mu|^2-\epsilon|\nu|^2+\delta|\tilde{u}|^2+\epsilon|\tilde{v}|^2]=0, \\
i\tilde{v}_t+2id\tilde{v}_x+\tilde{v}_{xx}+\tilde{v}[-\delta|\mu|^2-\epsilon|\nu|^2+\delta|\tilde{u}|^2+\epsilon|\tilde{v}|^2]=0.
\end{array} \label{(2.17)}
\end{equation}
Letting
\[\tilde{u}=ue^{i[(-\delta|\mu|^2-\epsilon|\nu|^2+c^2)t-cx]}, \]
\[\tilde{v}=ve^{i[(-\delta|\mu|^2-\epsilon|\nu|^2+d^2)t-dx]},\]
Eqs. \eqref{(2.17)} are then transformed into
\begin{equation}
\begin{array}{ll}
iu_t+u_{xx}+(\delta|u|^2+\epsilon|v|^2)u=0, \\
iv_t+v_{xx}+(\delta|u|^2+\epsilon|v|^2)v=0,
\end{array} \label{(2.18)}
\end{equation}
which has N-dark-dark soliton solutions as
\begin{equation}\label{(2.19)}
\begin{array}{ll}
u=\mu e^{i[cx+(\delta|\mu|^2+\epsilon|\nu|^2-c^2)t]}\frac{g_N}{f_N},
\\
v=\nu e^{i[dx+(\delta|\mu|^2+\epsilon|\nu|^2-d^2)t]}\frac{h_N}{f_N},
\end{array}
\end{equation}
with $f_N,g_N,h_N$ given by \eqref{(2.7)}. Finally, taking
$t\rightarrow-t$, Eqs. \eqref{(2.18)} become the generally coupled
NLS equations \eqref{(1.1)}. Hence we immediately have the following
theorem for solutions of Eq. (\ref{(1.1)}).

\begin{Theorem}
The N-dark-dark soliton solutions for the generally coupled NLS
equations \eqref{(1.1)} are
\begin{equation}
\begin{array}{ll}u=\mu e^{i[cx-(\delta|\mu|^2+\epsilon|\nu|^2-c^2)t]}\frac{G_N}{F_N},\\
v=\nu e^{i[dx-(\delta|\mu|^2+\epsilon|\nu|^2-d^2)t]}\frac{H_N}{F_N},
\end{array} \label{(2.9)}
\end{equation}
where
\begin{equation}
\begin{array}{ll}
F_N=\Big|\delta_{ij}+\frac{1}{p_i+\bar p_j}
e^{\theta_i+\bar\theta_j}\Big|_{N\times N},\\
G_N=\Big|\delta_{ij}-\frac{1}{p_i+\bar p_j}\frac{p_i-ic}{\bar
p_j+ic} e^{\theta_i+\bar\theta_j}\Big|_{N\times N},\\
H_N=\Big|\delta_{ij}-\frac{1}{p_i+\bar p_j}\frac{p_i-id}{\bar
p_j+id} e^{\theta_i+\bar\theta_j}\Big|_{N\times N},
\end{array} \label{(2.10)}
\end{equation}
\[\theta_j=p_jx-ip_j^2t+\theta_{j0},\]
$c,d$ are real constants, $\mu,\nu,p_j,\theta_{j0}$ are complex
constants, and these constants satisfy the following constraints
\begin{equation} \label{theorem2_constraints}
\frac{\delta|\mu|^2}{|p_j-ic|^2}+\frac{\epsilon|\nu|^2}{|p_j-id|^2}=-2,
~~~~~j=1,2,\cdots,N.
\end{equation}
\end{Theorem}
These solitons are dark-dark solitons, i.e., both $u$ and $v$
components are dark solitons, because it is easy to verify that
\begin{equation} \label{bc}
\begin{array}{ll}
u \to \mu e^{i[cx-(\delta|\mu|^2+\epsilon|\nu|^2-c^2)t +
\phi_{\pm}]}, \\
v \to \nu e^{i[dx-(\delta|\mu|^2+\epsilon|\nu|^2-d^2)t+\chi_\pm]},
\end{array}  \quad x\to \pm\infty,
\end{equation}
where $\phi_\pm$ and $\chi_\pm$ are phase constants. Thus the $u$
and $v$ solutions approach constant amplitudes $|\mu|$ and $|\nu|$
at large distances. When $\delta>0$ and $\epsilon>0$, which
correspond to self-focusing nonlinearities for both $u$ and $v$
components in Eqs. (\ref{(1.1)}), the constraints
(\ref{theorem2_constraints}) can not be satisfied, thus dark-dark
solitons can not exist as expected. When $\delta<0$ and
$\epsilon<0$, which correspond to self-defocusing nonlinearities for
both $u$ and $v$ components, dark-dark solitons can exist as Ref.
\cite{Ablowitz1} shows. A new phenomenon revealed by Theorem 2 is
that, when $\delta$ and $\epsilon$ have opposite signs, which
correspond to mixed focusing and defocusing nonlinearities in the
$u$ and $v$ equations, the constraints (\ref{theorem2_constraints})
can still be satisfied, hence dark-dark solitons can still exist.
This phenomenon will be demonstrated in more detail in the next
section. Interestingly, when $\delta$ and $\epsilon$ have opposite
signs, Eqs. (\ref{(1.1)}) also admit bright-bright solitons
\cite{Wang2010}. Thus Eqs. (\ref{(1.1)}) with opposite signs of
$\delta$ and $\epsilon$ are the rare equations which support both
dark-dark and bright-bright solitons.

The parameter constraints (\ref{theorem2_constraints}) can be solved
explicitly, so that solutions (\ref{(2.9)}) can be expressed in
terms of free parameters only. Let us write
\[p_j=a_j+ib_j, \]
where $a_j$ and $b_j$ are the real and imaginary parts of $p_j$.
Then Eq. (\ref{theorem2_constraints}) becomes
\begin{equation} \label{aj_equation}
\frac{\delta|\mu|^2}{a_j^2+(b_j-c)^2}+\frac{\epsilon|\nu|^2}{a_j^2+(b_j-d)^2}=-2.
\end{equation}
Solving this equation, we find that $a_j^2$ can be obtained
explicitly as
 \begin{eqnarray}  \label{aj_formula}
 a_j^2 & = & \frac{1}{2}\left\{-\left[(b_j-c)^2+(b_j-d)^2+\frac{1}{2}\delta |\mu|^2+\frac{1}{2}\epsilon
 |\nu|^2\right] \right.    \nonumber \\
 & & \hspace{0.5cm}  \left. \pm \sqrt{\left[(b_j-c)^2-(b_j-d)^2+\frac{1}{2}\delta |\mu|^2-\frac{1}{2}\epsilon
 |\nu|^2\right]^2+\delta\epsilon |\mu|^2|\nu|^2} \right\}.
 \end{eqnarray}
Here $\delta, \epsilon, \mu, \nu, c, d$ and $b_j$ are all free
parameters as long as the quantity under the square root of
(\ref{aj_formula}) as well as the whole right hand side of
(\ref{aj_formula}) are non-negative. If $a_j\le 0$, we will see that
the soliton solution (\ref{(2.9)}) would be singular. Thus in this
paper, we will always take $a_j>0$ to avoid this singularity.

We would like to make four remarks here. The first remark is on the
above derivation of dark solitons through KP-hierarchy reduction.
This derivation is non-trivial. To better understand it, we can
split it into two parts. One part is the reduction of the bilinear
equations (\ref{(2.6)}) of the generally coupled NLS equations
\eqref{(1.1)} from the KP-hierarchy equations (\ref{(2.1)}). The
other part is the reduction of the soliton solutions to the bilinear
equations (\ref{(2.6)}) from the $\tau$-solutions (\ref{(2.2)}) of
the KP-hierarchy equations (\ref{(2.1)}). In the first part, when we
impose on the $\tau$-functions the conjugation constraint [see
(\ref{conjugation_constraint_0})]
\begin{equation}  \label{conjugation_constraint}
\tau(k,l)=\overline{\tau(-k,-l)},
\end{equation}
and the linear constraint [see (\ref{(2.81)})]
\begin{equation} \label{linear_constraint}
(\delta|\mu|^2\partial_r+\epsilon|\nu|^2\partial_s)\tau(k,l)=-2\partial_x\tau(k,l),
\end{equation}
and set
\[ f=\tau(0,0), \quad g=\tau(1,0), \quad h=\tau(0, 1), \quad y=it,
\quad a=ic, \quad b=id,
\]
with $t, c, d$ being real, then one can readily verify that the
KP-hierarchy equations (\ref{(2.1)}) reduce to the bilinear
equations (\ref{(2.6)}) of the coupled NLS equations \eqref{(1.1)}.
In the second part, in order for the $\tau$-functions (\ref{(2.2)})
to satisfy the conjugation constraint
(\ref{conjugation_constraint}), it is sufficient to require [see
(\ref{conjugation_constraint_0})]
\begin{equation} \label{mji_ij}
m_{ji}(k,l)=\overline{m_{ij}(-k,-l)}.
\end{equation}
A sufficient condition for (\ref{mji_ij}) to hold is that
\begin{equation} \label{cqeta_cond}
c_{ij}=\delta_{ij}, \quad q_j=\bar p_j,  \quad \eta_j=\bar{\xi}_j,
\quad \eta_{j0}=\bar\xi_{j0},
\end{equation}
$x,r,s$ are real, and $y,a,b$ are pure imaginary. These conditions
are the same ones we imposed at the beginning of the proof of
Theorem 1. Under these conditions, the $\tau$-solutions
(\ref{(2.2)}) of the KP-hierarchy equations (\ref{(2.1)}) then
reduce to the solutions (\ref{(2.7)}) for the bilinear equations
(\ref{(2.6)}) of the coupled NLS equations \eqref{(1.1)}. In order
for the $\tau$-functions (\ref{(2.2)}) to satisfy the linear
constraint (\ref{linear_constraint}), by rewriting these
$\tau$-functions as (\ref{(2.800)}) and inserting them into this
linear constraint, we then get the parameter constraint
(\ref{constraint2}), which is equivalent to the parameter constraint
(\ref{theorem2_constraints}) in Theorem 1. This splitting of the
earlier derivation of dark solitons into these two parts helps to
clarify this derivation and make it more understandable.

The second remark is on the solution form (\ref{(2.9)}) of dark
solitons in the generally coupled NLS equations \eqref{(1.1)}. It is
known that the NLS equation of focusing type is a reduction of the
two-component KP hierarchy (see \cite{JM}, page 966 and 999), and
the NLS equation of defocusing type is a reduction of the
single-component KP hierarchy \cite{O}. It is also known that
solutions to the single-component KP hierarchy can be expressed as
single Wronskians \cite{Hirota,Freeman_Nimmo_1983,Nimmo_1989}, and
solutions to the two-component KP hierarchy can be expressed as
double Wronskians \cite{Freeman_1990}. Thus $N$-bright solitons in
the focusing NLS equation can be expressed as double Wronskians
\cite{Nimmo,Freeman}, and $N$-dark solitons in the defocusing NLS
equation can be expressed as single Wronskians \cite{O}. These
Wronskian solutions can also be expressed as Gram-type determinants
\cite{Hirota,MOS,Freeman_Nimmo_1983,Na,Ni}. For the vector
generalization \eqref{(1.1)} of the NLS equation, in order to obtain
its $N$-bright-soliton solutions, one should increase the number of
components, and take \eqref{(1.1)} as a reduction of the
three-component KP hierarchy. Thus $N$-bright solitons in
\eqref{(1.1)} can be expressed as three-component Wronskians (or the
corresponding Gram-type determinants \cite{Hirota}). But to obtain
$N$-dark solitons in Eqs. \eqref{(1.1)}, one should increase copies
of independent variables to $(r,k)$ and $(s,l)$ in the
single-component KP hierarchy [see Eqs. (\ref{(2.1)})], thus
$N$-dark solitons in Eqs. \eqref{(1.1)} can still be expressed as
single Wronskian (or the corresponding Gram determinant) as we have
done above.

The third remark we make is on comparison of the KP-hierarchy
reduction method and the inverse scattering method for deriving
dark-soliton solutions. As is well known, the inverse scattering
method is another way to derive soliton solutions. For bright
solitons, the inverse scattering method (or its modern
Riemann-Hilbert formulation) is a powerful way to derive such
solutions (see \cite{Zakharov_book, Shchesnovich_Yang} for
instance). Recently, bright-bright $N$-solitons in a very general
class of integrable coupled NLS equations were easily derived by
this method \cite{Wang2010}, and Eqs. \eqref{(1.1)} are special
cases of such general equations. But for dark solitons, the inverse
scattering method is more difficult due to non-vanishing boundary
conditions, which create branch cuts and other related intricacies
in the scattering process \cite{Faddeev_book}. In \cite{Ablowitz1},
the inverse scattering transform analysis was developed for the
defocusing Manakov equations [$\delta=\epsilon=-1$ in \eqref{(1.1)}]
with non-vanishing boundary conditions. But in their analysis, the
boundary conditions (\ref{bc}) were taken such that $c=d$ [see their
equation (2.3)] (actually $c=d=0$ was taken there, but the case of
$c=d\ne 0$ can be reduced to the case of $c=d=0$ through Galilean
transformation). When $c=d$, one can see from our general formula
(\ref{(2.9)}) that $u$ and $v$ are simply proportional to each
other, thus their inverse scattering analysis could only obtain
degenerate dark-dark solitons which are reducible to scalar dark
solitons in the defocusing NLS equation. In order to derive the more
general dark-dark solitons (\ref{(2.9)}) with $c\ne d$, the inverse
scattering method would be even more complicated than that in
\cite{Ablowitz1}. Comparatively, the KP-hierarchy reduction method
we used above is free of these difficulties, and is thus a simpler
method for deriving dark-soliton solutions.

Our last remark is on dark solitons in an even more general coupled
NLS equations
\begin{equation} \label{more_general}
\begin{array}{ll}
iu_{t}=u_{xx}+ \left( \delta |u|^{2}+\epsilon |v|^{2}+\gamma u\bar
v+\bar \gamma \bar uv \right)u, \\ iv_{t}=v_{xx}+ \left(
\delta|u|^{2}+\epsilon|v|^{2}+\gamma u\bar v+\bar \gamma \bar uv
\right)v,
\end{array}
\end{equation}
where $\delta, \epsilon$ are real constants as in \eqref{(1.1)}, and
$\gamma$ is a complex constant. If $\gamma=0$, (\ref{more_general})
reduces to \eqref{(1.1)}). This more general coupled NLS system
(\ref{more_general}) is also integrable. Its Lax pair as well as
$N$-bright-bright solitons are given in \cite{Wang2010}. To explore
dark-dark solitons in this system, we look for solutions with the
following large-distance asymptotics [as in (\ref{(2.9)})]
\begin{equation}
\left\{
\begin{array}{ll}u \to \mu e^{i[cx-\omega t]}, \\
v\to \nu e^{i[dx-\kappa t]},
\end{array} \right. \quad x \to -\infty, \label{large_x}
\end{equation}
where $\mu, \nu$ are non-zero complex constants, and $c, d, \omega,
\kappa$ are real constants. Inserting this asymptotic solution into
(\ref{more_general}), we see that due to the $\gamma$-terms, Eqs.
(\ref{more_general}) can hold only if $c=d$, and $\omega=\kappa$.
Based on the previous solutions (\ref{(2.9)}), this would imply that
the $u$ and $v$ components of dark-dark solitons in the general
system (\ref{more_general}) must be proportional to each other, thus
are equivalent to scalar dark solitons in the defocusing NLS
equation. Except these trivial dark-dark solitons, Eqs.
(\ref{more_general}) do not admit other dark-dark solitons of the
form (\ref{large_x}) when $\gamma\ne 0$. This is a dramatic
difference between the cases of $\gamma=0$ and $\gamma\ne 0$ in Eqs.
(\ref{more_general}). Whether the general system
(\ref{more_general}) admits dark-dark solitons with background
asymptotics different from (\ref{large_x}) is still unclear.

\section{Dynamics of dark solitons}

In what follows, we investigate the dynamics of single-dark-soliton
and two-dark-soliton solutions in the generally coupled NLS
equations \eqref{(1.1)}. In the analysis of these solutions,
$\delta$ and $\epsilon$ will be treated as  arbitrary parameters. In
the illustrations of solutions in the figures, we will pick
\begin{equation}
\delta=1, \quad \epsilon=-1,
\end{equation}
which correspond to mixed focusing and defocusing nonlinearities.
The reason for this choice is that dark solitons under such mixed
nonlinearities have never been studied before. We will show that
under these mixed nonlinearities, some novel phenomena (such as
existence of two-dark-soliton bound states) would arise. Soliton
dynamics under other $\delta$ and $\epsilon$ values, such as in the
defocusing Manakov equations where $\delta=\epsilon=-1$, would also
be briefly discussed when appropriate.

\subsection{Single dark solitons}

\begin{figure}[h]
\begin{center}
\includegraphics[width=0.75\textwidth]{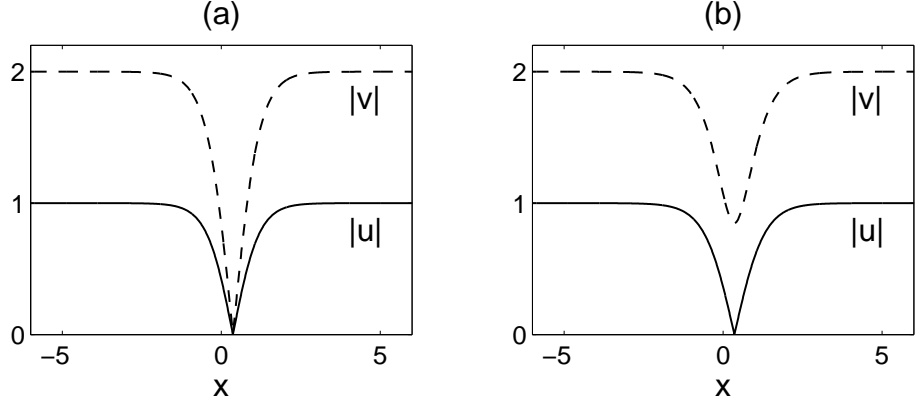}
\end{center}
\caption{Single dark-dark solitons in Eqs. (\ref{(1.1)}) with
$\delta=1, \epsilon=-1$: (a) a degenerate soliton with parameters
(\ref{parameter_S1}); (b) a non-degenerate soliton with parameters
(\ref{parameter_S2}).} \label{fig_1soliton}
\end{figure}

In order to get single dark solitons in Eqs. \eqref{(1.1)}, we set
$N = 1$ in the formula \eqref{(2.9)}. After simple algebra, these
single dark solitons can be written as
\begin{equation}
\begin{array}{ll}
u=\frac{1}{2}\mu e^{i[cx-(\delta|\mu|^2+\epsilon|\nu|^2-c^2)t]}
\left[1+y_1+(y_1-1)\tanh(\frac{\theta_1+\bar\theta_1+\rho_1}{2})\right],
\end{array} \label{(3.1)}
\end{equation}
\begin{equation}
\begin{array}{ll}
v=\frac{1}{2}\nu e^{i[dx-(\delta|\mu|^2+\epsilon|\nu|^2-d^2)t]}
\left[1+z_1+(z_1-1)\tanh(\frac{\theta_1+\bar\theta_1+\rho_1}{2})\right],
\end{array} \label{(3.2)}
\end{equation}
where
\[\theta_1=p_1x-ip_1^2t+\theta_{10}, \quad e^{\rho_1}=1/(p_1+\bar p_1),\]
\[y_1=(ic-p_1)/(ic+\bar p_1), \quad z_1=(id-p_1)/(id+\bar p_1),\]
and $\mu,\nu,p_1,\theta_{10}$ are complex constants satisfying
\begin{equation} \label{constraint_1soliton}
\frac{\delta|\mu|^2}{|p_1-ic|^2}+\frac{\epsilon|\nu|^2}{|p_1-id|^2}=-2,
\end{equation}
or equivalently, $a_1$ is given by formula (\ref{aj_formula}), where
$p_1=a_1+ib_1$. This soliton would be singular if $p_1+\bar p_1\le
0$, i.e., $a_1\le 0$. Thus we will require $a_1>0$ below to avoid
singular solutions. It is easy to see that the intensity functions
$|u|$ and $|v|$ of these dark solitons move at velocity
$-2\hspace{0.06cm}b_1$. In addition, they approach constant
amplitudes $|\mu|$ and $|\nu|$ respectively as $x\to \pm \infty$. As
$x$ varies from $-\infty$ to $+\infty$, the phases of the $u$ and
$v$ components acquire shifts in the amount of $2\phi_1$ and
$2\chi_1$, where
\begin{equation}
y_1=e^{2i\phi_1}, \quad z_1=e^{2i\chi_1},
\end{equation}
i.e., $2\phi_1$ and $2\chi_1$ are the phases of constants $y_1$ and
$z_1$ respectively. Without loss of generality, we restrict $-\pi <
2\phi_1, 2\chi_1 \le \pi$, i.e., $-\pi/2< \phi_1, \chi_1\le \pi/2$.
At the center of the soliton where $\theta_1+\bar\theta_1+\rho_1=0$,
intensities of the two components are
\begin{equation}
|u|_{center}=|\mu|\cos\phi_1, \quad |v|_{center}=|\nu|\cos\chi_1.
\end{equation}
These center intensities are lower than the background intensities
$|\mu|$ and $|\nu|$, thus these solitons are dark solitons. Notice
that the center intensities of the $u$ and $v$ solutions are
controlled by their respective phase shifts $2\phi_1$ and $2\chi_1$,
thus these phase shifts dictate how ``dark" the center is. This
general single dark-dark soliton (\ref{(3.1)})-(\ref{(3.2)}) has
been derived for the defocusing Manakov model before by the Hirota
method in \cite{RL,Sheppard_Kivshar_1997}. In particular, a
parameter constraint similar to (\ref{constraint_1soliton}) was
given in \cite{RL}. If $c=d$, then $y_1=z_1$, hence $\phi_1=\chi_1$.
In this case, the $u$ and $v$ components are proportional to each
other, and have the same degrees of darkness at the center. This
soliton is equivalent to a scalar dark soliton in the defocusing NLS
equation, thus is degenerate. It is noted that the single-dark-dark
soliton derived in \cite{Ablowitz1} [see Eq. (5.8) there]
corresponds to this degenerate type of dark-dark solitons. To
illustrate, we take
\begin{equation} \label{parameter_S1}
\mu=1, \quad \nu=2,, \quad c=d=0, \quad p_1=\sqrt{1.5}, \quad
\theta_{10}=0,
\end{equation}
which satisfy the constraint (\ref{constraint_1soliton}).
Intensities of the solution (\ref{(3.1)})-(\ref{(3.2)}) are
displayed in Fig. \ref{fig_1soliton}(a). This soliton is stationary,
and both its $u$ and $v$ components are black (with zero intensity)
at the soliton center.

Non-degenerate single-dark-dark-solitons in Eqs.  \eqref{(1.1)},
however, are such that $c\ne d$. The $u$ and $v$ components in these
solitons are not proportional to each other, thus are not reducible
to scalar single dark solitons in the defocusing NLS equation. Since
$c\ne d$, $y_1\ne z_1$, thus $\phi_1\ne \chi_1$. This means that the
$u$ and $v$ components in these non-degenerate solitons have
different degrees of darkness at its center. To illustrate, we take
\begin{equation} \label{parameter_S2}
\mu=1, \quad \nu=2, \quad c=0, \quad d=0.5, \quad p_1=1.0679,
\end{equation}
which also satisfies the constraint (\ref{constraint_1soliton}).
Here the $p_1$ value is obtained from the formula (\ref{aj_formula})
with the plus sign and $b_1=0$. Intensities of this soliton are
displayed in Fig. \ref{fig_1soliton}(b). This soliton is also
stationary. At its center, the $u$ component is black, but the $v$
component is only gray. This type of non-degenerate single dark-dark
solitons in the coupled NLS system (\ref{(1.1)}) has not been
obtained before (to our knowledge).

In the defocusing Manakov equations where $\delta=\epsilon=-1$,
their degenerate and non-degenerate single dark solitons
qualitatively resemble those shown in Fig. \ref{fig_1soliton}, and
are thus not shown.

\subsection{Collision of two dark solitons}

Two-dark-soliton solutions in system \eqref{(1.1)} correspond to
$N=2$ in the general formula \eqref{(2.9)}. In this case, we have

\begin{equation}
u=\mu e^{i[cx-(\delta|\mu|^2+\epsilon|\nu|^2-c^2)t]}\frac{G_2(x,
t)}{F_2(x, t)}, \label{(3.5)}
\end{equation}
\begin{equation}
v=\nu e^{i[dx-(\delta|\mu|^2+\epsilon|\nu|^2-d^2)t]}\frac{H_2(x,
t)}{F_2(x, t)}, \label{(3.6)}
\end{equation}
where
\begin{eqnarray}
F_2(x,t) & = &
1+e^{\theta_1+\bar\theta_1+\rho_1}+e^{\theta_2+\bar\theta_2+\rho_2}
+r e^{\theta_1+\bar\theta_1+\theta_2+\bar\theta_2+\rho_1+\rho_2},
\\
G_2(x,t) & = &
1+y_1e^{\theta_1+\bar\theta_1+\rho_1}+y_2e^{\theta_2+\bar\theta_2+\rho_2}+
ry_1y_2e^{\theta_1+\bar\theta_1+\theta_2+\bar\theta_2+\rho_1+\rho_2},
\\
H_2(x,t) & = &
1+z_1e^{\theta_1+\bar\theta_1+\rho_1}+z_2e^{\theta_2+\bar\theta_2+\rho_2}+
rz_1z_2e^{\theta_1+\bar\theta_1+\theta_2+\bar\theta_2+\rho_1+\rho_2},
\end{eqnarray}
\begin{equation} \label{rhoj}
\theta_j=p_j x-ip_j^2t+\theta_{j0}, \quad e^{\rho_j} =1/(p_j+\bar
p_j),
\end{equation}
\begin{equation}  \label{yjzj}
y_j=(ic-p_j)/(ic+\bar p_j), \quad z_j=(id-p_j)/(id+\bar p_j),
\end{equation}
\begin{equation} \label{rdef}
r=1-(p_1+\bar p_1)(p_2+\bar p_2)/|p_1+\bar p_2|^2,
\end{equation}
and $\mu, \nu, p_1, p_2, \theta_{10},\theta_{20}$ are complex
constants satisfying the constraint (\ref{theorem2_constraints})
with $j=1, 2$, or equivalently, $a_j$ is given by the formula
(\ref{aj_formula}), where $p_j=a_j+ib_j$.

\begin{figure}[h]
\begin{center}
\includegraphics[width=0.8\textwidth]{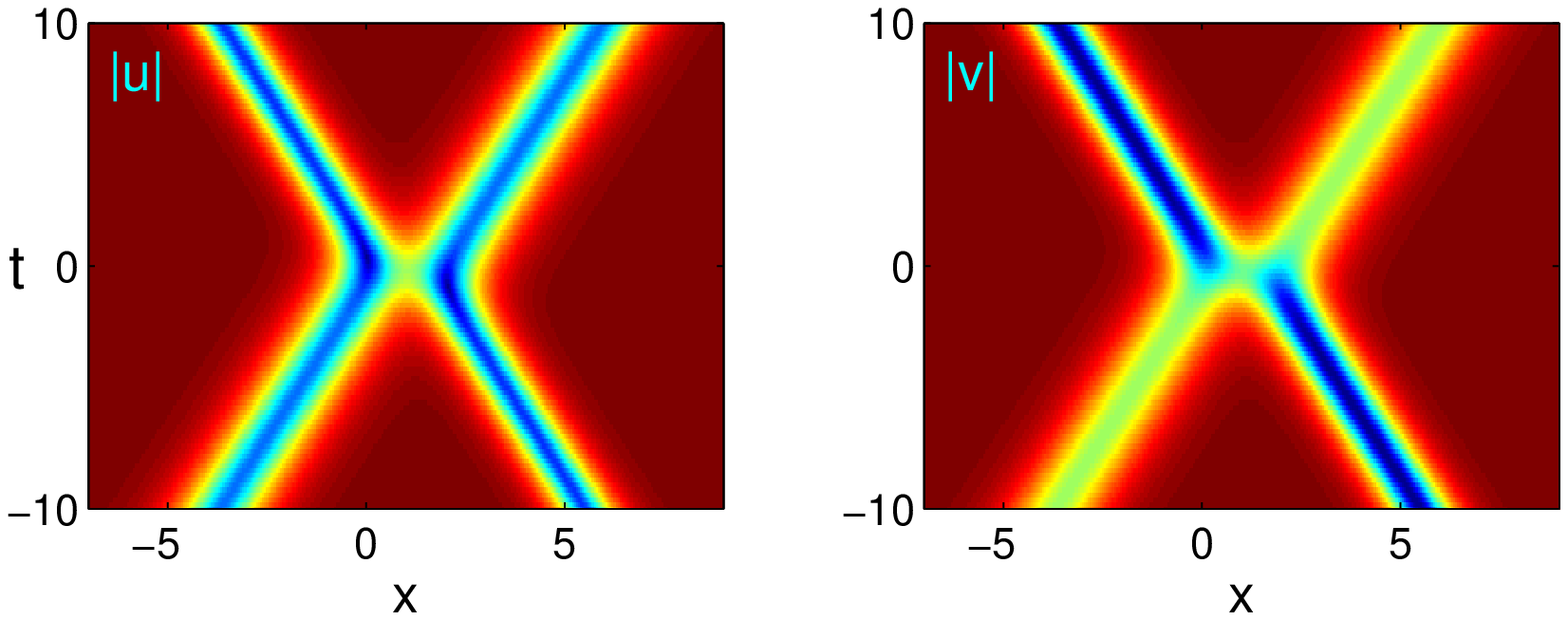}

\vspace{0.6cm}
\includegraphics[width=0.8\textwidth]{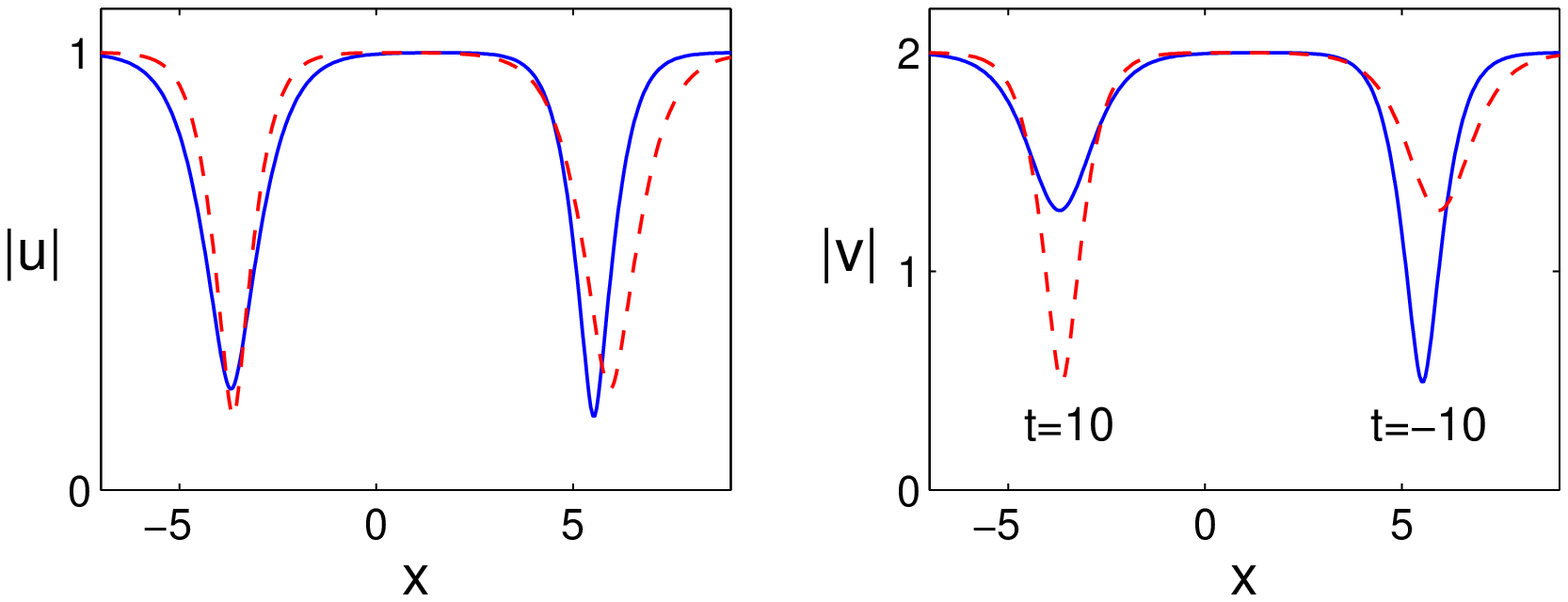}
\end{center}
\caption{Collision of two dark-dark solitons in Eqs. \eqref{(1.1)}
with $\delta=1, \epsilon=-1$ and parameters (\ref{ic_fig1}). The
upper row shows the $(x, t)$ evolution, and the lower row shows the
intensity profiles before and after collision: $t=-10$ (solid);
$t=10$ (dashed). } \label{fig_collision}
\end{figure}

In generic cases where $\mbox{Im}(p_1) \ne \mbox{Im}(p_2)$, these
solutions describe the collision of two dark-dark solitons. To
demonstrate these collisions, we take parameters
\begin{equation} \label{ic_fig1}
\mu=1, \hspace{0.15cm} \nu=2,  \hspace{0.15cm} c=0, \hspace{0.15cm}
d=0.5, \hspace{0.15cm} p_1=0.8426 - 0.2i, \hspace{0.15cm} p_2=1.1801
+ 0.2i, \hspace{0.15cm} \theta_{10}=\theta_{20}=0.
\end{equation}
Here the real parts of $p_1$ and $p_2$ are obtained from the formula
(\ref{aj_formula}) with the plus sign. The corresponding two
dark-dark soliton solution \eqref{(3.5)}-\eqref{(3.6)} is shown in
Fig. \ref{fig_collision}. We can see that after collision, the two
dark solitons pass through each other without any change of shape
and velocity in either of its two components. Hence the degrees of
darkness in each soliton do not change after collision, which means
that there is no energy transfer from one component to the other
inside each soliton after collision. In addition, there is no energy
transfer from one soliton to the other after collision either. This
complete transmission of dark solitons' energy in both its two
components after collision occurs not only for $\delta=1$ and
$\epsilon=-1$ as in Fig. \ref{fig_collision}, but also for all other
$\delta$ and $\epsilon$ values. Thus it is a common phenomenon of
the generally coupled NLS system (\ref{(1.1)}). For instance, it
also happens in the defocusing Manakov equations where
$\delta=\epsilon=-1$.

This complete transmission of dark-dark solitons' energy in both its
two components is a remarkable phenomenon, because it is in stark
contrast with collisions of bright-bright solitons in the same
coupled NLS system (\ref{(1.1)}). Indeed, for bright-bright solitons
in the focusing Manakov system (with $\delta=\epsilon=1$),
polarization rotations take place after collision, hence energy has
transferred from one component to the other in each soliton
\cite{Manakov}. For bright-bright solitons in the more general
coupled NLS system (\ref{more_general}) (such as $\delta=1$ and
$\epsilon=-1$ above), energy can also transfer from one soliton to
another after collision \cite{Wang2010}. Thus collisions between
bright-bright solitons and between dark-dark solitons in the coupled
NLS system (\ref{(1.1)}) are distinctly different.

The reason for this complete energy transmission in all components
in dark-soliton collisions is that the intensity profile of each
dark-dark soliton is completely characterized by the background
parameters $\mu, \nu, c, d$ and the soliton parameter $p_j$ [see
Eqs. (\ref{(3.1)})-(\ref{(3.2)})]. These background parameters are
the same for both colliding solitons, and clearly do not change
before and after collision. The soliton parameter $p_j$ corresponds
to the spectral discrete eigenvalue in the inverse scattering
transform method, and is a constant of motion throughout collision.
Consequently, the intensity profile of each dark-dark soliton (in
both $u$ and $v$ components) can not change before and after
collision. This property indicates that dark solitons are more
robust than bright solitons with regard to collision. The positions
of dark solitons do shift after collision though, as can be seen
clearly in Fig. \ref{fig_collision}. This position shift is always
toward the soliton's moving direction, which is the same as
collisions of bright solitons in the NLS equation
\cite{Zakharov_Shabat}.

\section{Dark-dark-soliton bound states}

In studies of dark solitons, multi-dark-soliton bound states is an
interesting subject. In the defocusing NLS equation, two dark
solitons repel each other, thus can not form a bound state
\cite{Kivshar_1993}. In the defocusing Manakov model,
multi-bright-dark-soliton bound states were reported in
\cite{Sheppard_Kivshar_1997}. Some of those bound states are
stationary, while the others are not. So far,
multi-dark-dark-soliton bound states have never been reported in
integrable systems. In a non-integrable system, namely, the
second-harmonic-generation (SHG) system, two-dark-dark-soliton bound
states do exist, as was reported in \cite{Buryak_Kivshar_1995}. In
this system, single dark-dark solitons with non-monotonic tails
exist. When two such dark-dark solitons weakly overlap with each
other and interact, their non-monotonic tails create local minima in
the effective interaction potential, hence the two dark-dark
solitons can form stationary bound states. In addition, some of
these bound states are stable \cite{Buryak_Kivshar_1995}.

In this section, we show that in the generally coupled NLS system
(\ref{(1.1)}), when both $\delta$ and $\epsilon$ are negative, i.e.,
all nonlinearities are defocusing (the defocusing Manakov model),
multi-dark-dark-soliton bound states can not exist. But for mixed
focusing and defocusing nonlinearities, where $\delta$ and
$\epsilon$ have opposite signs, two-dark-dark-soliton bound states
do exist and are stationary. To our knowledge, this is the first
report of multi-dark-dark-soliton bound states in integrable
systems. Properties and physical origins of these stationary bound
states in the mixed-nonlinearity model (\ref{(1.1)}) are quite
different from the stationary bright-dark-soliton bound states in
the defocusing Manakov model \cite{Buryak_Kivshar_1995} and
stationary dark-dark-soliton bound states in the non-integrable SHG
model \cite{Buryak_Kivshar_1995}, as we will explain later in this
section.

To obtain dark-dark-soliton bound states, the two dark solitons in
the solution (\ref{(3.5)})-(\ref{(3.6)}) should have the same
velocity, i.e., $\mbox{Im}(p_1)=\mbox{Im}(p_2)$ (or $b_1=b_2$), so
that the two constituent dark solitons can stay together for all
times. In order for this to happen, two different (positive) values
$a_1$ and $a_2$ from Eq. (\ref{aj_equation}) must exist for the same
values of $b_1=b_2$. When $\delta$ and $\epsilon$ are both negative,
where the nonlinearities are all defocusing, this is not possible.
The reason is that when $\delta<0$ and $\epsilon<0$, the function on
the left side of Eq. (\ref{aj_equation}) is an increasing function
of $a_j^2$. Thus for this function to reach the value level of $-2$
on the right side of Eq. (\ref{aj_equation}), there is at most one
$a_j^2$ solution, hence at most one positive $a_j$ value. This means
that when nonlinearities are all defocusing (i.e., the defocusing
Manakov model with $\delta=\epsilon=-1$), there are no
multi-dark-dark-soliton bound states. However, when $\delta$ and
$\epsilon$ have opposite signs, where focusing and defocusing
nonlinearities are mixed, the function on the left side of Eq.
(\ref{aj_equation}) may become non-monotone in $a_j^2$, hence it
becomes possible for Eq. (\ref{aj_equation}) to admit two different
positive values $a_1$ and $a_2$ for the same values of $b_1=b_2$
(see below). In the formula (\ref{aj_formula}), these different
$a_1$ and $a_2$ values correspond to the plus and minus signs
respectively. In this case, two-dark-dark-soliton bound states would
exist, and this is a new phenomenon in the coupled NLS equations
(\ref{(1.1)}) under mixed focusing and defocusing nonlinearities.
Physically, these results on bound states in Eqs. (\ref{(1.1)}) can
be heuristically understood as follows. We know that in the scalar
defocusing NLS equation, two dark solitons repel each other. In the
coupled NLS system (\ref{(1.1)}), if $\delta$ and $\epsilon$ are
both negative, all nonlinearities are defocusing, hence two
dark-dark solitons still repel each other, and no bound states can
be formed. However, if $\delta$ and $\epsilon$ have opposite signs,
parts of the nonlinear terms are focusing, and the other parts
defocusing. While the defocusing terms repel two dark solitons, the
focusing terms do just the opposite, which is to attract two dark
solitons. Thus, when these repulsive and attractive forces balance
each other, two dark-dark solitons then can form a stationary bound
state. This physical mechanism for the existence of
dark-dark-soliton bound states is quite different from that in the
SHG model \cite{Buryak_Kivshar_1995} (see earlier text).

Next we examine these two-dark-dark-soliton bound states in more
detail. Through Galilean transformation (i.e., in the moving
coordinate system with this common velocity), this common velocity
can be reduced to zero. Hence $p_1$ and $p_2$ become real
parameters. In this case, it is easy to see that this bound state
becomes
\begin{equation}
\begin{array}{ll}u=\mu e^{i[cx-(\delta|\mu|^2+\epsilon|\nu|^2-c^2)t]}\frac{G_2(x)}{F_2(x)},\\
v=\nu
e^{i[dx-(\delta|\mu|^2+\epsilon|\nu|^2-d^2)t]}\frac{H_2(x)}{F_2(x)},
\end{array}
\end{equation}
where
\begin{eqnarray}
F_2(x) & = & 1+e^{2p_1x+2\alpha_1+\rho_1}+e^{2p_2x+2\alpha_2+\rho_2}
+r e^{2p_1x+2p_2x+2\alpha_1+2\alpha_2+\rho_1+\rho_2},
\\
G_2(x) & = &
1+y_1e^{2p_1x+2\alpha_1+\rho_1}+y_2e^{2p_2x+2\alpha_2+\rho_2}+
ry_1y_2e^{2p_1x+2p_2x+2\alpha_1+2\alpha_2+\rho_1+\rho_2},
\\
H_2(x) & = &
1+z_1e^{2p_1x+2\alpha_1+\rho_1}+z_2e^{2p_2x+2\alpha_2+\rho_2}+
rz_1z_2e^{2p_1x+2p_2x+2\alpha_1+2\alpha_2+\rho_1+\rho_2},
\end{eqnarray}
$\alpha_j=\mbox{Re}(\theta_{j0})$, and $\rho_j, y_j, z_j, r$ are as
given in Eqs. (\ref{rhoj})-(\ref{rdef}). Notice that functions $F_2,
G_2$ and $H_2$ are time-independent, thus this bound state is
actually stationary. This is analogous to certain
bright-dark-soliton bound states in the defocusing Manakov model
\cite{Sheppard_Kivshar_1997} and dark-dark-soliton bound states in
the SHG model \cite{Buryak_Kivshar_1995}. An important feature of
these present bound states is that, as $x$ moves from $-\infty$ to
$+\infty$, these states acquire non-zero phase shifts. Indeed, it is
easy to see from the above solution formula that the phase shifts of
the $u$ and $v$ components are
\begin{equation} \label{phase_shift}
 u\mbox{-phase shift}=2\phi_1+2\phi_2, \quad  v\mbox{-phase
 shift}=2\chi_1+2\chi_2,
\end{equation}
where $2\phi_j$ and $2\chi_j$ are the phases of $y_j$ and $z_j$
respectively. In other words, the total phase shifts of the bound
state are equal to the sum of the individual phase shifts of the two
constituent dark solitons, which are non-zero in general. This
contrasts stationary bright-dark-soliton bound states in the
defocusing Manakov model \cite{Sheppard_Kivshar_1997} and
dark-dark-soliton bound states in the SHG model
\cite{Buryak_Kivshar_1995}, where phase shifts of the dark
components across the soliton are all zero.

\begin{figure}[h]
\begin{center}
\includegraphics[width=0.85\textwidth]{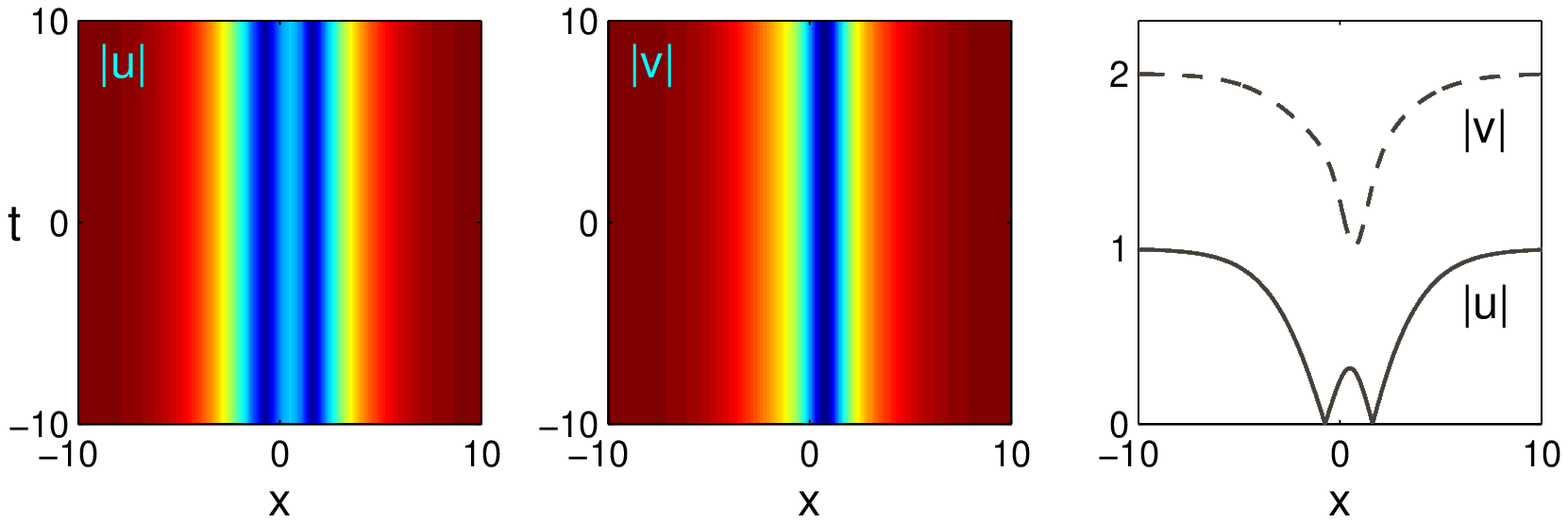}

\vspace{0.5cm}
\includegraphics[width=0.85\textwidth]{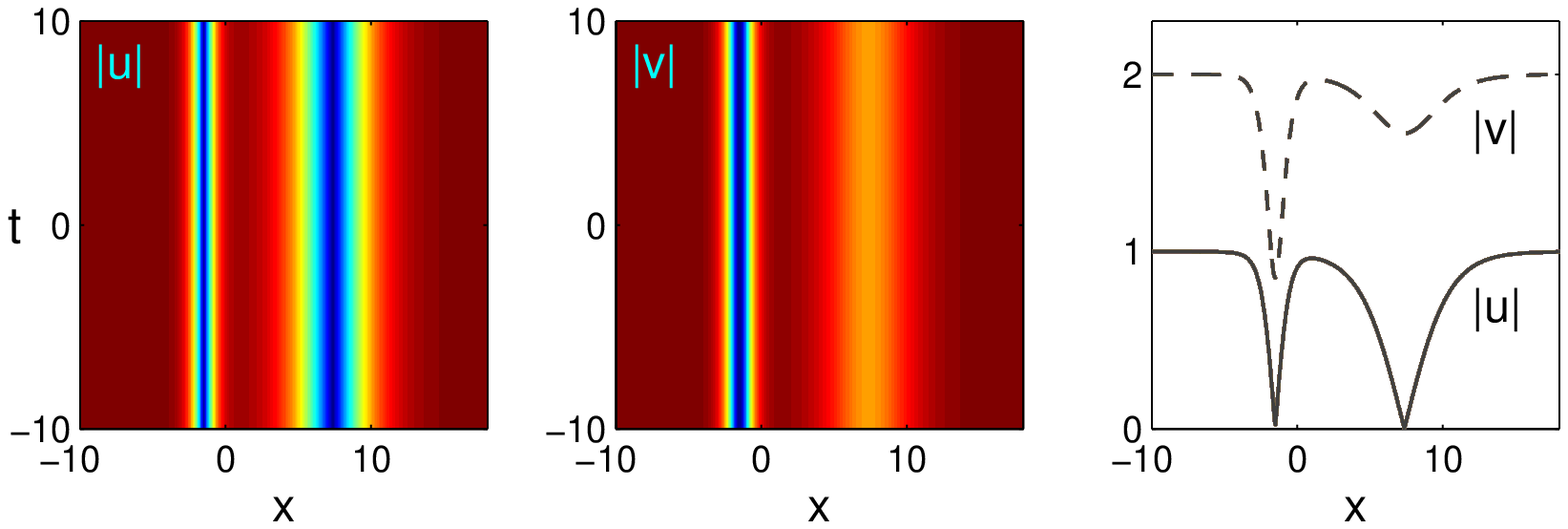}
\end{center}
\caption{Two examples of two-dark-soliton bound states in Eqs.
\eqref{(1.1)} with $\delta=1, \epsilon=-1$. Upper row: bound state
with parameters (\ref{ic_bound_state}); lower row: bound state with
parameters (\ref{ic_bound_state}) except that
$\alpha_1=-\alpha_2=2$. The left two panels show $(x, t)$ evolution
of $|u|$ and $|v|$ components, and the right panel shows the
stationary intensity profiles. } \label{fig_bound_state}
\end{figure}

To demonstrate these stationary two-dark-soliton bound states, we
take parameters
\begin{equation} \label{ic_bound_state}
\mu=1, \hspace{0.15cm} \nu=2,  \hspace{0.15cm} c=0, \hspace{0.15cm}
d=0.5, \hspace{0.15cm} p_1=1.0679,  \hspace{0.15cm} p_2=0.3311,
\hspace{0.15cm} \alpha_1=\alpha_2=0.
\end{equation}
Here $p_1$ and $p_2$ are obtained from the formula
(\ref{aj_formula}) with $b_1=b_2=0$. The corresponding bound state
is displayed in Fig. \ref{fig_bound_state} (upper row). In this
bound state, the $u$-component is double-dipped (i.e., has a double
hole), signifying this is a two-soliton bound state, while the
$v$-component is single-dipped. By adjusting $\alpha_1$ and
$\alpha_2$ values, we can obtain bound states where both $u$ and $v$
components are double-dipped. For instance, when we take
$\alpha_1=-\alpha_2=2$ instead of zero in (\ref{ic_bound_state}), we
get such a bound state which is shown in the lower row of Fig.
\ref{fig_bound_state}. For both bound states, the total phase shift
of the $u$-component is zero, and the total phase shift of the
$v$-component is 3.4355, as can be calculated from formula
(\ref{phase_shift}).

From the above analytical formulae and Fig. \ref{fig_bound_state},
we can see that these stationary two-dark-soliton bound states have
six free parameters, $\mu, \nu, c, d, \alpha_1$ and $\alpha_2$ [the
positive $p_1$ and $p_2$ values are determined from formula
(\ref{aj_formula}) by setting $b_1=b_2=0$]. The first four
parameters characterize the background intensities and phase
gradients, while the parameters $\alpha_1$ and $\alpha_2$ control
the positions of the two dark solitons.

The above dark-soliton bound states in the integrable coupled NLS
system (\ref{(1.1)}) possess properties which are very different
from those in dark-soliton bound states in the non-integrable SHG
model \cite{Buryak_Kivshar_1995}. First, the bound states in the SHG
model are formed by identical dark solitons (see Fig. 4 in
\cite{Buryak_Kivshar_1995}), but the bound states in the coupled NLS
system are formed by different dark solitons since $p_1\ne p_2$ (see
lower row of Fig. \ref{fig_bound_state} in this paper). Second, the
bound states in the SHG model have zero phase shifts from one end to
the other, but the phase shifts of bound states in the coupled NLS
system (\ref{(1.1)}) are non-zero in general [see Eqs.
(\ref{phase_shift})]. Thirdly, the bound states in the SHG model
have non-zero binding energy, hence can be stable against
perturbations \cite{Buryak_Kivshar_1995}. But the bound states in
the present coupled NLS system have zero binding energy. Thus under
perturbations, the two constituent dark solitons in these bound
states generically will split apart, analogously to bright-soliton
bound states in the focusing NLS equation.

At this point, one may wonder if three- and higher-dark-dark-soliton
bound states exist in the coupled NLS system (\ref{(1.1)}). It turns
out that such bound states can not exist. The reason is that, in a
bound state, velocities of all constituent solitons must be the
same, i.e., all $b_j$ [i.e., $\mbox{Im}(p_j)$] must be the same. In
order for three- and higher-dark-soliton bound states to exist,
formula (\ref{aj_formula}) must give at least three distinct
positive solutions $a_j$ for the same $b_j$ value. This is clearly
impossible, since formula (\ref{aj_formula}) can give at most two
distinct positive $a_j$ values when the plus and minus signs are
taken. Consequently, three- and higher-dark-dark-soliton bound
states can not exist in Eqs. (\ref{(1.1)}). Note that in the
defocusing Manakov model, non-stationary three and higher
bright-dark-soliton bound states exist, but stationary three and
higher bright-dark-soliton bound states do not
\cite{Sheppard_Kivshar_1997}; while in the non-integrable SHG
system, stationary three and higher dark-dark-soliton bound states
do exist \cite{Buryak_Kivshar_1995}.

\section{Summary and discussion}

In this paper, we have investigated dark-dark solitons in the
integrable generally coupled NLS system (\ref{(1.1)}). By reducing
the Gram-type solution of the KP hierarchy, we derived the general
$N$-dark-dark solitons in this system. We showed that the dark-dark
solitons derived previously in the literature are only degenerate
cases of these general soliton solutions. We have also shown that
when these solitons collide with each other, energies in both
components of the solitons completely transmit through. This
behavior contrasts bright-bright solitons in this system, where
polarization rotation and soliton reflection can occur after
collision. In addition, we have shown that when focusing and
defocusing nonlinearities are mixed, two dark-dark solitons can form
a stationary bound state. These results will be useful for many
physical subjects such as nonlinear optics, water waves and
Bose-Einstein condensates, where the coupled NLS equations often
arise.

The dark-dark solitons obtained in this paper for the generally
coupled NLS system (\ref{(1.1)}) are useful for other purposes as
well. For instance, it is known that from dark solitons of the
defocusing NLS equation, one can obtain homoclinic solutions of the
focusing NLS equation through simple variable transformations
\cite{Ablowitz_homoclinic}. Thus, from these dark-dark solitons in
this paper, we can obtain homoclinic solutions to these generally
coupled NLS equations (\ref{(1.1)}). Since solutions near homoclinic
orbits often exhibit chaotic dynamics \cite{Ablowitz_homoclinic},
the homoclinic solutions for the generally coupled NLS equations
(\ref{(1.1)}) then can serve as the starting point to understand
chaotic behaviors in these systems.

Lastly, we would like to mention that $N$-bright-bright and
$N$-bright-dark solitons in the coupled NLS equations (\ref{(1.1)})
can also be obtained by the KP-hierarchy reduction method. But those
reductions will be different from the ones in this paper for
dark-dark solitons, and will be left for future studies.

\vskip .2cm \noindent{\bf Acknowledgments}

We thank Dr. Xingbiao Hu for helpful discussions. The work of Y.O.
was partly supported by JSPS Grant-in-Aid for Scientific Research
(B-19340031, S-19104002). The work of D.S.W. was supported by China
Postdoctoral Science Foundation. The work of J.Y. was supported in
part by the (U.S.) Air Force Office of Scientific Research under
grant USAF 9550-09-1-0228 and the National Science Foundation under
grant DMS-0908167.


\begin{thebibliography}{30}
\bibitem{Benney} D.J. Benney and A.C. Newell, Nonlinear wave envelopes, J. Math. Phys. 46, 133 (1967).
\bibitem{Agrawal_book} G.P. Agrawal, \emph{Nonlinear Fiber Optics}, (Academic Press, San
Diego, 1989).
\bibitem{Hasegawa_book}
A. Hasegawa and Y. Kodama, \emph{Solitons in Optical
Communications,} (Clarendon, Oxford, 1995).
\bibitem{Ablowitz_Segur} M.J. Ablowitz and H. Segur, \emph{Solitons and
the Inverse Scattering Transform} (SIAM, Philadelphia, 1981).
\bibitem{Dalfovo_1999}
F. Dalfovo, S. Giorgini, L. P. Pitaevskii, and S. Stringari, Theory
of Bose-Einstein condensation in trapped gases, Rev. Mod. Phys. 71,
463 (1999).
\bibitem{Ho_Shenoy}T.-L. Ho and V. B. Shenoy, Hartree-Fock theory for double
condensates, Phys. Rev. Lett. 77, 3276 (1996).
\bibitem{Pu_Bigelow_1}H. Pu and N. P.
Bigelow, Properties of two-species Bose condensates, Phys. Rev.
Lett. 80, 1130 (1998).
\bibitem{Pu_Bigelow_2}H. Pu and N. P.
Bigelow, Collective excitations, metastability, and nonlinear
response of a trapped two-species Bose-Einstein condensate, Phys.
Rev. Lett. 80, 1134 (1998).
\bibitem{Goldstein_Meystre}I. Goldstein and P. Meystre, A priori definition of maximal CP nonconservation,
Phys. Rev. A 55, 2935 (1997).
\bibitem{Roskes}
G.J. Roskes, Some nonlinear multiphase interactions, Stud. Appl.
Math. 55, 231-238 (1976).
\bibitem{Menyuk_1987} C.R. Menyuk, Nonlinear pulse
propagation in birefringent optical fibers, IEEE J. Quantum
Electron. 23, 174 (1987).
\bibitem{Zakharov_Shabat}
V.E. Zakharov and A.B. Shabat, Exact theory of two-dimensional self-
focusing and one-dimensional self-modulation of waves in nonlinear
media, Zh. E'ksp. Teor. Fiz. 61, 118 (1971) [Sov. Phys. JETP 34, 62
(1972)].
\bibitem{Faddeev_book}
L.D. Faddeev and L.A. Takhtadjan, \emph{Hamiltonian Methods in the
Theory of Solitons} (Springer Verlag, Berlin, 1987).
\bibitem{Manakov} S.V. Manakov, On the theory of two-dimensional stationary self-focusing of electromagnetic waves,
Zh. Eksp. Teor. Fiz 65, 1392 (1973) [Sov. Phys. JETP 38, 248-253
(1974)].
\bibitem{Zakharov1982}V.E. Zakharov and E.I. Schulman, To the integrability of the system of two coupled nonlinear
Schr${\rm \ddot{o}}$dinger equations Physica 4D, 270 (1982).
\bibitem{Wang2010} D.S. Wang, D. Zhang and J. Yang, Integrable properties of the
general coupled nonlinear Schr${\rm \ddot{o}}$dinger equations, J.
Math. Phys. 51, 023510 (2010).
\bibitem{Sheppard_Kivshar_1997} A.P. Sheppard and Y.S. Kivshar,
Polarized dark solitons in isotropic Kerr media, Phys. Rev. E 55,
4773 (1997).
\bibitem{RL} R. Radhakrishnan and M. Lakshmanan,
Bright and dark soliton solutions to coupled nonlinear Schr\"odinger
equations, J. Phys. A: Math. Gen. 28, 2683-2692 (1995).
\bibitem{Ablowitz1}B. Prinari, M. J. Ablowitz, and G. Biondini, Inverse scattering transform for the vector nonlinear
Schr\"{o}dinger equation with nonvanishing boundary conditions, J.
Math. Phys. 47, 063508 (2006).
\bibitem{KLTA} T. Kanna, M. Lakshmanan, P. Tchofo Dinda, and N.
Akhmediev, Soliton collisions with shape change by intensity
redistribution in mixed coupled nonlinear Schr\"odinger equations,
Phys. Rev. E 73, 026604 (2006).
\bibitem{VKL_2008}
M. Vijayajayanthi, T. Kanna, and M. Lakshmanan, Bright-dark solitons
and their collisions in mixed N-coupled nonlinear Schr\"odinger
equations, Phys. Rev. A 77, 013820 (2008).
\bibitem{DKJM} E. Date, M. Kashiwara, M. Jimbo and T. Miwa,
Transformation groups for soliton equations,
\emph{Nonlinear Integrable Systems---Classical Theory and Quantum Theory},
39-119 (World Scientific, Singapore, 1983).
\bibitem{T} K. Takasaki,
Geometry of universal Grassmann manifold from algebraic point of view,
Rev. Math. Phys. 1, 1-46 (1989).
\bibitem{Date} E. Date, M. Jimbo, M. Kashiwara and T. Miwa, Operator approach to the Kadomtsev-Petviashvili
equation-transformation groups for soliton equations III, J. Phys.
Soc. Jpn. 50, 3806 (1981).
\bibitem{O} Y. Ohta,
Wronskian solutions of soliton equations, RIMS Kokyuroku 684, 1-17
(1989). [In Japanese]
\bibitem{Hirota}R. Hirota, \emph{The Direct Method in Soliton Theory} (Cambridge University
Press, Cambridge, 2004).
\bibitem{MOS} S. Miyake, Y. Ohta and J. Satsuma,
A representation of solutions for the KP hierarchy and its algebraic
structure, J. Phys. Soc. Jpn. 59, 48-55 (1990).
\bibitem{OHTI} Y. Ohta, R. Hirota, S. Tsujimoto and T. Imai,
Casorati and discrete Gram type determinant representations of
solutions to the discrete KP hierarchy, J. Phys. Soc. Jpn. 62,
1872-1886 (1993).
\bibitem{JM} M. Jimbo and T. Miwa,
Solitons and infinite-dimensional Lie algebras, Publ. Res. Inst.
Math. Sci. 19, 943-1001 (1983).
\bibitem{DJM} E. Date, M. Jimbo and T. Miwa,
Method for generating discrete soliton equations. II, J. Phys. Soc.
Jpn. 51, 4125-4131 (1982).
\bibitem{Freeman_Nimmo_1983} N.C. Freeman and J.J.C. Nimmo,
Soliton solutions of the Korteweg de Vries and the
Kadomtsev-Petviashvili equations: the Wronskian technique. Proc. R.
Soc. A 389 319 (1983).
\bibitem{Nimmo_1989}
J.J.C. Nimmo, Wronskian determinants, the KP hierarchy and
supersymmetric polynomials. J. Phys. A: Math. Gen. 22, pp. 3213-3221
(1989).
\bibitem{Freeman_1990}
N. C. Freeman, C. R. Gilson, and J. J. Nimmo, Two-component KP
hierarchy and the classical Boussinesq equation, J. Phys. A: Math.
Gen. 23, 4793--4803 (1990).

\bibitem{Nimmo} J.J.C.  Nimmo, A bilinear B\"acklund transformation for the
nonlinear Schr\"odinger equation. Phys. Lett. A, 99, 279--280
(1983).
\bibitem{Freeman}
N. C. Freeman, Soliton solutions of nonlinear evolution equations.
IMA J. Appl. Math. 32, 125--145 (1984).

\bibitem{Na} A. Nakamura,
A bilinear $N$-soliton formula for the KP equation,
J. Phys. Soc. Jpn. 58, 412-422 (1989).
\bibitem{Ni} J.J.C. Nimmo,
Darboux transformations for a two-dimensional Zakharov-Shabat/AKNS
spectral problem,
Inv. Prob. 8, 219-243 (1992).
\bibitem{Zakharov_book}
V. E. Zakharov, S. V. Manakov, S. P. Novikov, and L. P. Pitaevskii,
\emph{The Theory of Solitons: The Inverse Scattering Method}
(Consultants Bureau, New York, 1984).
\bibitem{Shchesnovich_Yang}
V.S. Shchesnovich and J. Yang, General soliton matrices in the
Riemann-Hilbert problem for integrable nonlinear equations, J. Math.
Phys. 44, 4604 (2003).
\bibitem{Kivshar_1993} Y.S. Kivshar, Dark solitons in nonlinear
optics, IEEE J. Quantum Electron. 29, 250 (1993).
\bibitem{Buryak_Kivshar_1995}
A.V. Buryak and Y.S. Kivshar, Twin-hole dark solitons, Phys. Rev. A
51, R41 (1995).

\bibitem{Ablowitz_homoclinic} M.J. Ablowitz and B.M. Herbst,
On homoclinic structure and numerically induced chaos for the
nonlinear Schr\"odinger equation, SIAM J. Appl. Math. 50, pp.
339-351 (1990).

\end{thebibliography}
\end{document}